\documentclass[11pt]{article}
\usepackage{graphicx}
\newcommand{\pr}{\\[1.3mm]\noindent}
\newcommand{\vB}{\mbox{\boldmath $B$}}
\newcommand{\vv}{\mbox{\boldmath $v$}}
\newcommand{\na}{\nabla}
\newcommand{\pa}{\partial}
\newcommand{\tauc}{\tau_{\rm c}}

\begin{document}

\title{\bf An analysis of the fluctuations of the geomagnetic dipole}

\author{K. Brendel$\,^{\rm a\,},\,$ J. Kuipers$\,^{\rm a\,},\,$ 
        G.T. Barkema$\,^{\rm a\,,\,}$\thanks{e-mail: 
        g.t.barkema@phys.uu.nl}{}\ \ \  
        and 
        P. Hoyng$\,^{\rm b\,,\,}$\thanks{e-mail: p.hoyng@sron.nl 
        (corresponding author)}
\\[5.mm]
$^{\rm a\,}$\small Theoretical Physics, Utrecht University, Leuvenlaan 4, 
3584 CE Utrecht,\\ The Netherlands; \\
$^{\rm b\,}$\small SRON Netherlands Institute for Space Research, 
Sorbonnelaan 2, 3584 CA Utrecht,\\ The Netherlands\\}

\date{\today}

\maketitle


\begin{abstract} 
The time evolution of the strength of the Earth's virtual axial dipole 
moment (VADM) is analyzed by relating it to the Fokker-Planck equation,
which describes a random walk with VADM-dependent drift and diffusion 
coefficients. We demonstrate first that our method is able to retrieve
the correct shape of the drift and diffusion coefficients from a time
series generated by a test model. Analysis of the Sint-2000 data shows
that the geomagnetic dipole mode has a linear growth time of 
$20^{+13}_{-7}$ kyr, and that the nonlinear quenching of the growth rate 
follows a quadratic function of the type $[1-(x/x_0)^2]$. On theoretical 
grounds, the diffusive motion of the VADM is expected to be driven by 
multiplicative noise, and the corresponding diffusion coefficient to scale 
quadratically with dipole strength. However, analysis of the Sint-2000 
VADM data reveals a diffusion which depends only very weakly on the dipole 
strength. This may indicate that the magnetic field quenches the amplitude 
of the turbulent velocity in the Earth's outer core.
\end{abstract}




\section{Introduction}
The strength of the geomagnetic dipole moment shows a considerable time 
variability, about 25\% r.m.s. of the mean, over the course of thousands of 
years. Occasionally, the variability is so large that the sign of the dipole 
moment changes. These reversals happen roughly once per $(2-3)\times 10^5$ 
yr (Merrill et al., 1996). The geomagnetic field is the result of inductive 
processes in the Earth's liquid metallic outer core. Helical convection 
amplifies the magnetic field and balances resistive decay. Several groups 
have confirmed this idea with the help of numerical simulations (Glatzmaier 
and Roberts, 1995; Kuang and Bloxham, 1997; Christensen et al., 1999). A 
suitable measure of the geomagnetic dipole is the Virtual Axial Dipole Moment 
(VADM), of which several records have been published, e.g. by Guyodo and Valet 
(1999) and Valet et al. (2005). Since the dipole moment is the result of many 
processes taking place in the convecting metallic outer core that interact 
with each other in a complicated way, it makes sense to try to describe the 
time evolution of the VADM over long time scales as a stochastic process. 


Before entering into details we recall that statistical modelling of the 
geomagnetic field has a long history. Constable and Parker (1988) were the 
first to give a complete characterization of the statistical properties of the 
geomagnetic field in terms of its spherical harmonic expansion coefficients. 
The distribution of the axial dipole was found to be symmetric and bi-modal, 
consisting of two gaussians shifted to the peak position of the two polarity 
states. They also showed that the expansion coefficients of the non-dipole 
field may, after appropriate scaling, be regarded as statistically independent 
samples of one single normal distribution with zero mean. This GGP (giant 
gaussian process) approach as it is now generally referred to, permitted 
computation of the average of any field-related quantity. Hulot and Le Mou\"el 
(1994) have extended the GGP approach by considering the evolution of the 
statistical properties with time, and Bouligand et al. (2005) have tested the 
GGP modelling technique on hydromagnetic geodynamo simulations.


\begin{figure}[t]
\centerline{\includegraphics[width=12.cm]{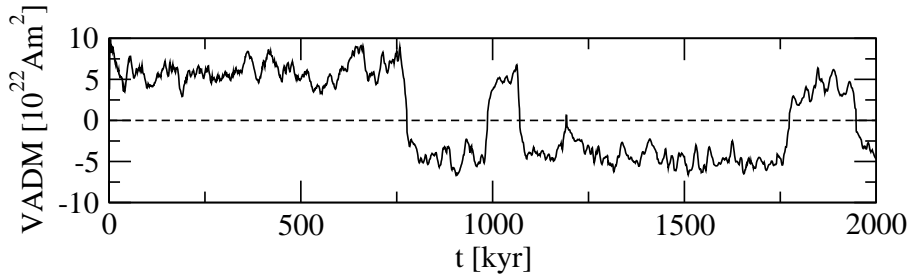}}
\caption{\small The Sint-2000 VADM data of Valet et al. (2005), a time series 
of 2000 unsigned VADM values spaced in time by 1000 year, covering the past 2 
Myr history. We have inserted a sign flip at the times of known reversals, 
see text.}  
\label{fig:sint2000}
\end{figure}


Returning to the time evolution of geomagnetic dipole as a stochastic process,
consider a stochastic equation of the type
\begin{equation}
\dot x= v(x)+F(x)L(t)\ .
\label{eq:langevin1}
\end{equation}
The function $v(x)$ has the dimension of a velocity and represents the 
effective growth rate of $x$, sometimes called the drift velocity. The 
fluctuations are embodied in the term $F(x)L(t)$ and they induce an additional 
diffusive motion of $x$.\footnote{The noise is called additive if $F$ is 
constant, otherwise it is referred to as multiplicative noise.} 
Here $L(t)$ is a stationary random function with zero mean and a short 
correlation time $\tauc$:
\begin{equation}
\langle L(t)\rangle=0,\quad \langle L(t)L(t-\tau)\rangle=
L_{\rm r.m.s.}^2\tauc\,\delta(\tau)\ .
\label{eq:langevin2}
\end{equation}
A short correlation time means that the duration $\tauc$ of the memory of 
$L(t)$ is much shorter than all other time scales in the process. Under these 
circumstances the autocorrelation function of $L(t)$ behaves as a 
$\delta$-function of time. The probability distribution $\rho(x,t)$ of $x(t)$ 
determined by Eq.~(\ref{eq:langevin1}) obeys the Fokker-Planck equation (Van 
Kampen, 1992; Gardiner, 1990):$\,$\footnote{Provided $v\tauc\ll x$; this 
particular form of Eq.~(\ref{eq:FP1}) requires in addition that $dD/dx\ll v$.} 
\begin{equation}
\frac{\partial \rho}{\partial t} =
  - \frac{\partial}{\partial x} (v\rho)
  + \frac{1}{2} \frac{\partial^2}{\partial x^2} \left( D \rho \right)\ .
\label{eq:FP1}
\end{equation}
Here $t$ is time, and $v$ is again the effective growth rate of $x$. The 
diffusion coefficient is equal to 
\begin{equation}
D\,\simeq\,2F^2\int_0^\infty\langle L(t)L(t-\tau)\rangle\,d\tau\,
\simeq\,F^2L_{\rm r.m.s.}^2\tauc\ . 
\label{eq:defD}
\end{equation}
%

%
%

The Fokker-Planck equation is a simple and versatile tool for modelling the 
dynamics of a stochastic process. That is to say, the statistical properties 
of a wide variety of different stochastic processes can be described by the 
Fokker-Planck equation (\ref{eq:FP1}). Hoyng et al. (2002) have shown that for 
theoretically plausible functions $v(x)$ and $D(x)$ the amplitude distribution 
of the Sint-800 data (Guyodo and Valet, 1999) is very well predicted by 
Eq.~(\ref{eq:FP1}).

The purpose of this paper is twofold. We investigate whether the Sint-2000 
VADM time series (Valet et al., 2005) can indeed be described by a 
Fokker-Planck equation (\ref{eq:FP1}). Secondly, we derive the dependence 
of the effective growth rate $v$ and the diffusion coefficient $D$ on the 
magnitude $x$ of the VADM without making any prior assumption on the 
functional form. In doing so we are able to measure the linear growth rate 
of the dipole mode and its nonlinear quenching from the data. Likewise, the 
diffusion coefficient $D(x)$ provides information on the convective flows in 
the outer core. This marks the difference between our approach and that of 
the GGP: we do not stop at giving a statistical desciption of the multipole 
coefficients of the geomagnetic field, but we extract information immediately 
related to the physics of the geomagnetic dipole. 
 
After a brief discussion of the Sint-2000 data in Section 2, we develop in 
Section 3 a technique for extracting the functions $v(x)$ and $D(x)$ from a 
time series. Next, in Section 4, we validate the method with the help of an 
artificial VADM time series generated by a simple model to see how well we 
can retrieve the $v(x)$ and $D(x)$ that were used to generate the series. In 
Section 5 we apply the method the Sint-2000 VADM data (Valet et al., 2005) 
and we discuss the implications of our findings for the geodynamo. A summary 
and our conclusions appear in Section 6. 


\begin{figure}[t]
\centerline{\includegraphics[width=8.cm]{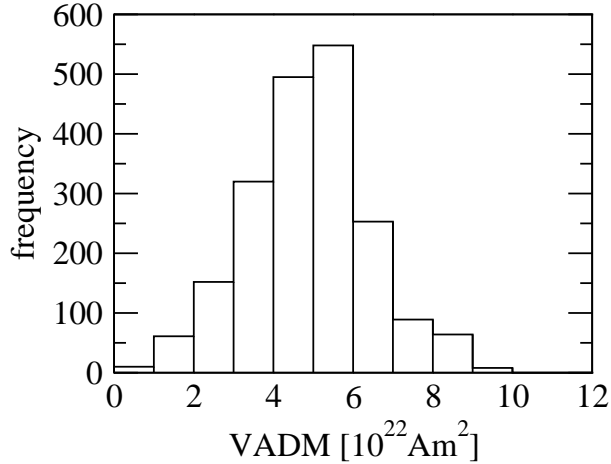}}
\caption{\small The amplitude distribution of the unsigned Sint-2000 data.
In principle the distribution is symmetric with respect to ${\rm VADM}=0$, 
and has a characteristic double-hump structure with small but nonzero 
probability at ${\rm VADM}=0$.
\label{fig:sint2000ampl}}
\end{figure}


\section{Sint-2000 data}
The Sint-2000 data comprises a time series of 2000 unsigned VADM values 
spaced by 1000 year, covering the past 2 Myr history of the geomagnetic 
dipole. The positions of the reversals are indicated in Figure~2 of Valet 
et al. (2005), where they show up as local minima in the VADM record. To 
obtain a VADM time series with sign we have inserted a sign change between 
those locations. The result is shown in Figure~\ref{fig:sint2000}. In 
doing so we may miss some of the fine structure in the reversal time 
profile. However, in view of the considerable intrinsic uncertainties in 
the data [see Figure~2 of Valet et al. (2005)], similar and larger 
ambiguities apply to the whole VADM time series. The amplitude distribution 
of the unsigned VADM data is shown in Figure~\ref{fig:sint2000ampl}.
                                                      

\section{Method}
We start with the discretized version of the Fokker-Planck equation.
Discretization of space and time in Eq.~\ref{eq:FP1} leads to
\begin{eqnarray}
\frac{\rho_i(t + \Delta t)\,-\,\rho_i(t)}{\Delta t}&=&
  -\;\frac{v_{i+1}\rho_{i+1}(t) - v_{i-1} \rho_{i-1}(t)}{2 \Delta x}
\nonumber \\[2.mm]
&&+\;\frac{D_{i+1}\rho_{i+1}(t) + D_{i-1}\rho_{i-1}(t) - 2 D_i\rho_i(t)}
  {2(\Delta x)^2}\ .
\end{eqnarray}
We may rewrite this equation in matrix form as
\begin{equation}
\rho_i(t+\Delta t) = \sum_j \left( \delta_{ij} + \Delta t M_{ij} \right)
\rho_j(t)\ ,
\label{eq:matrix}
\end{equation}
where is $M$ a tridiagonal matrix with elements
\begin{eqnarray}
M_{i,i-1} &=& \frac{v_{i-1}}{2\Delta x} +\frac{D_{i-1}}{2(\Delta x)^2}\ ;
\nonumber \\
M_{i,i} &=& -\,\frac{D_i}{(\Delta x)^2}\ ;
\nonumber \\
M_{i,i+1} &=& -\,\frac{v_{i+1}}{2\Delta x} +\frac{D_{i+1}}{2(\Delta x)^2}\ .
\label{eq:Mii}
\end{eqnarray}
By solving for $v_i$ and $D_i$ we obtain
\begin{eqnarray}
v_i &=& \left( M_{i+1,i} - M_{i-1,i} \right) \Delta x\ ;
\nonumber \\[1.mm]
D_i &=& \left( M_{i+1,i} + M_{i-1,i} \right) (\Delta x)^2\ .
\label{eq:viDi}
\end{eqnarray}
These expressions will be used to infer $v_i$ and $D_i$ once the matrix
$M$ has been determined.

Each column of $M$ adds up to zero, $M_{i-1,i}+M_{i,i}+M_{i+1,i}=0$, so
that there are two free parameters per spatial interval $i$, exactly as
many as the parameters $v_i$ and $D_i$ in the Fokker-Planck equation. An
important consequence of the zero column sum is that Eq.~(\ref{eq:matrix})
is norm-conserving,
\begin{equation}
\sum_i\rho_i(t+\Delta t) = \sum_i \rho_i(t)\ .
\label{eq:noco}
\end{equation}
In the limit $\Delta t\rightarrow 0$ Eq.~(\ref{eq:matrix}) becomes
\begin{equation}
\frac{d\rho_i}{dt}=\sum_j M_{ij}\rho_j\ .
\label{eq:dvgl}
\end{equation}
The object of this paper is to extract the effective position-dependent 
(= VADM-dependent) velocity and diffusion coefficient from a time series, 
in this case of the strength of the Earth's magnetic dipole moment. 
To this end we construct from the data the matrix $T$ whose elements 
$T_{ij}$ contain the transition probabilities for a system in position 
$j$ at some time $t$ to move to position $i$ at a later time $t+\tau$. 
We begin by counting the number of times $N_{ij}$ that the system is 
located in position $j$ at some time $t$ and in position $i$ at time 
$t+\tau$. The required matrix elements are then equal to $T_{ij}=
\alpha_j\cdot N_{ij}$ and have a statistical error $\sigma_{ij}=
\alpha_j\cdot N_{ij}{}^{1/2}$. The normalization coefficients $\alpha_j$ 
are fixed by the requirement that the columns of $T_{ij}$ should add up 
to unity, $\sum_iT_{ij}=1$. The time lag $\tau$, finally, must be chosen 
comparable to, or larger than the correlation time of the randomly 
fluctuating part of the system, but small in comparison to the time 
scale on which the data changes systematically.

Our assumption is that the process is Markovian and therefore can be
described by Eq.~(\ref{eq:dvgl}), from which it follows that 
$\rho_i(t+\tau)=\sum_j\exp(\tau M)_{ij}\rho_j(t)$. The theoretical 
transition matrix $\tilde{T}$ is therefore 
\begin{equation}
\tilde{T}\,=\,\exp(\tau M)\ ,
\label{eq:tildeT}
\end{equation}
and our goal is now to find a tridiagonal matrix $M$ such that $\tilde{T}$ 
closely resembles $T$. The matrix $M$ has approximately $3n$ degrees of 
freedom (ignoring boundary effects), of which $n$ can be eliminated by norm 
conservation (columns add up to zero). To find the remaining $2n$ degrees 
of freedom we minimize the function
\begin{equation}
\sum_{i,j}\left(\frac{T_{ij}-\exp(\tau M)_{ij}}{\sigma_{ij}}\right)^2\ .
\label{eq:minim}
\end{equation}
We could follow an alternative approach, by using the stationary 
distribution $p_i\equiv\rho_i(\infty)$ which we may find by binning the 
data as in Figure~\ref{fig:sint2000ampl}. Since the stationary distribution 
should obey Eq.~(\ref{eq:dvgl}), we have $\sum_j M_{ij}p_j=0$. This 
relation can be used to eliminate another $n$ degrees of freedom in $M$, 
after which we find the remaining $n$ by minimizing the function 
(\ref{eq:minim}). But we opted for fitting $2n$ degrees of freedom and to 
use $\sum_j M_{ij}p_j=0$, or equivalently $\sum_j \tilde{T}_{ij}p_j=p_i\,$, 
as a consistency check on our computations.


\begin{figure}[t]
\centerline{\includegraphics[width=6.cm]{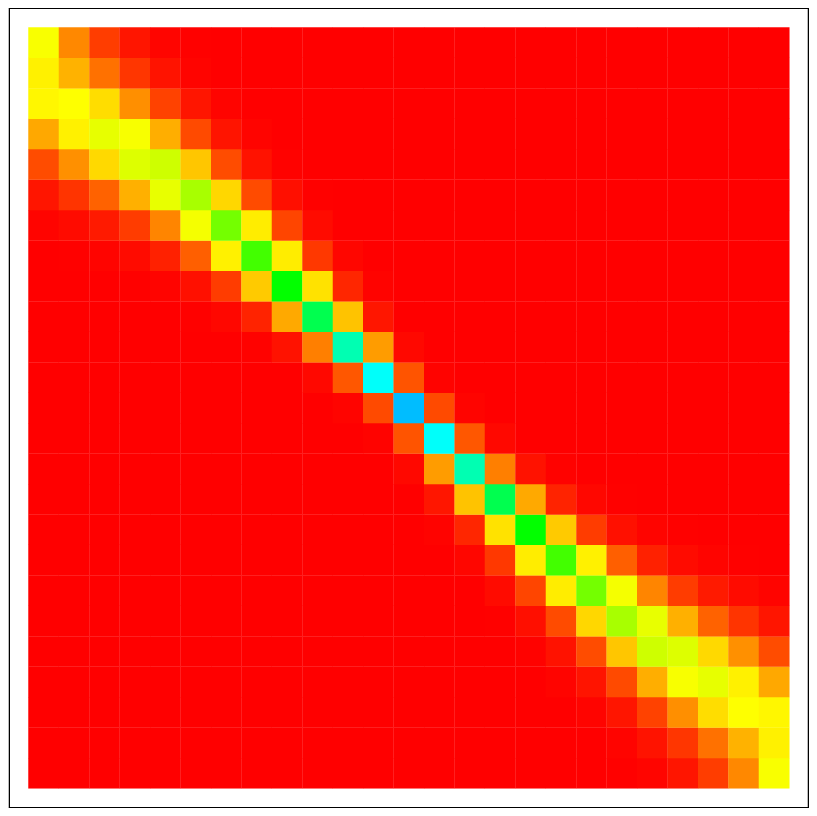}
            \includegraphics[width=6.cm]{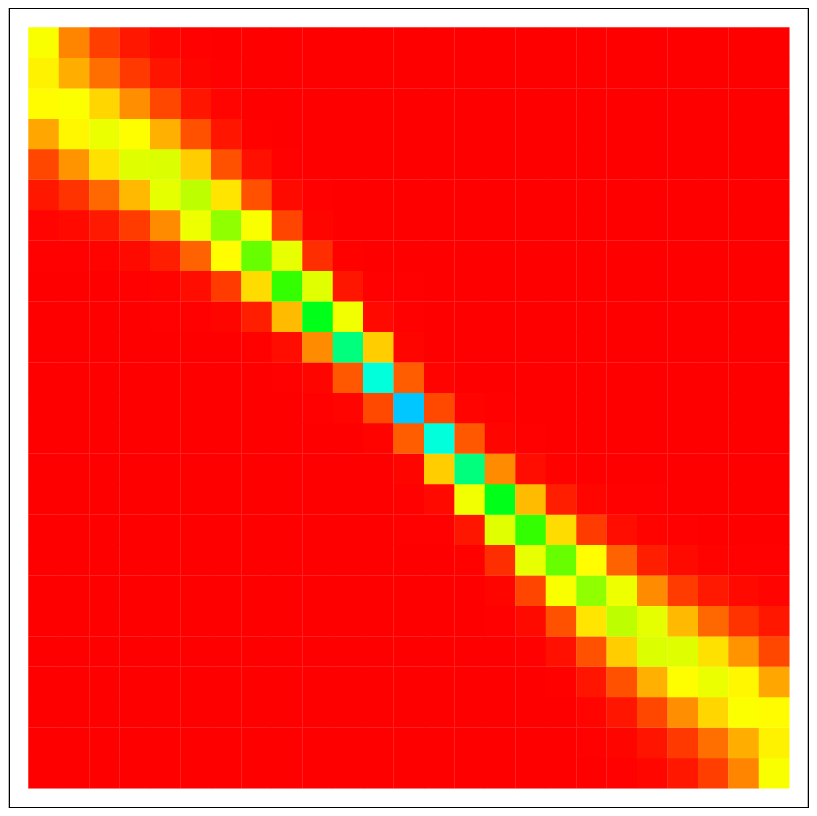}}
\centerline{\includegraphics[width=21.4cm]{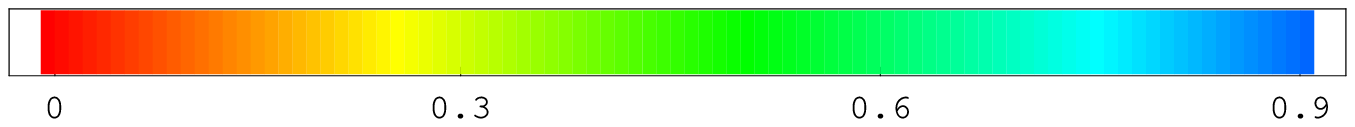}}
\caption{\small The left panel is the transition matrix $T$ as obtained 
from simulation data of the HD model with $\tau=0.007$. Right panel: the 
approximate transition matrix $\tilde{T}=\exp(\tau M)$, where $M$ is a 
tridiagonal matrix, see text. The upper left corner corresponds to $(-2,-2)$, 
the lower right to $(2,2)$. The bin size is $0.16\times 0.16$. The matrix 
elements obey $0\leq T_{ij}<1$ and $\sum_i T_{ij}=1$.
\label{fig:figsim}}
\end{figure}


\section{Validation with the HD model}
First, we test the approach outlined in the previous section on data generated 
with the model of Hoyng and Duistermaat (2004).\footnote{Henceforth referred 
to as `HD' or `HD model'.} 
This is a time series $x(t)$ of VADMs measured in units of the equilibrium 
value, so $x=1$ corresponds to the nonlinear equilibrium value of the VADM. 
The series comprises $5\times 10^6$ data points with a time spacing of 
$0.001$, and by construction this is also the correlation time.\footnote{The 
time resolution of this series is a factor 10 higher than that of the series 
used in Fig. 2 of HD, but the other parameters are the same ($a=2$, $c=5$, and 
$D=0.4$).}
Time is measured in units of the linear growth time of the dipole mode so that 
the series is about 50 Myr long in real time. We discretize the strength 
$x(t)$ of the VADM into 25 bins of width 0.16 in dimensionless units, and we 
construct a histogram of all sets $\{x(t), x(t+\tau)\}$ employing a time lag 
of $\tau=0.007$. We then exploit the fact that there is no sign preference, 
that is, for a given realisation $x(t)$ the series $-x(t)$ is an equally 
likely realisation. Accordingly, we add to the histogram all sets 
$\{-x(t),-x(t+\tau)\}$. We follow the procedure of the previous section, and
the resulting effective transition matrix $T$ is plotted in 
Figure~\ref{fig:figsim}, left panel. Note that the blue matrix elements near 
the centre have a relatively large value but not a more accurate one: matrix 
elements near the centre of the figure are determined by small $x(t)$ 
associated with reversals and these are rare. The most accurate elements 
correspond therefore to the equilibrium value $x=\pm 1$, and are located in 
the wings near $(1,1)$ and $(-1,-1)$ in Figure~\ref{fig:figsim} (left panel).


\begin{figure}[t]
\centerline{\includegraphics[width=9.5cm]{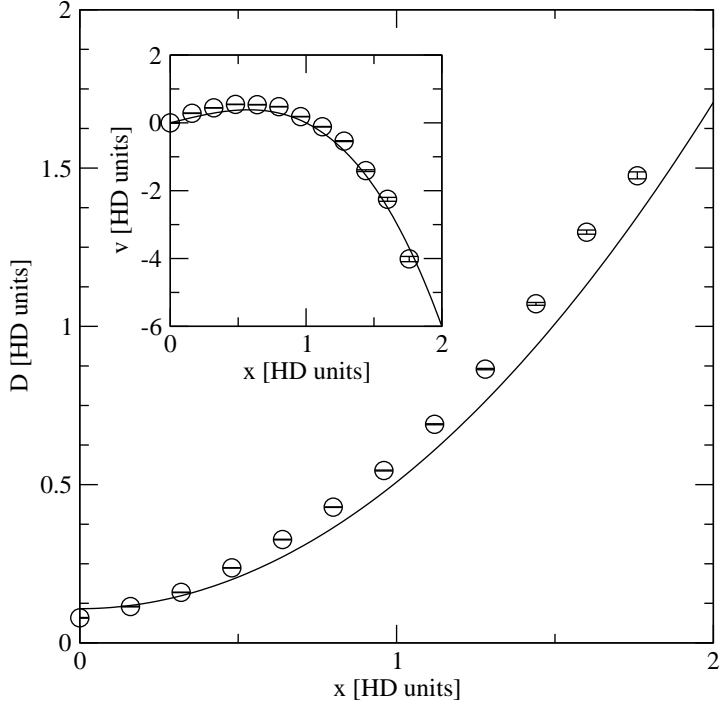}}
\caption{\small The diffusion coefficients $D_i$ as a function of $x$, 
obtained by fitting the simulation data of Hoyng and Duistermaat (2004) 
to the Fokker-Planck equation. The drawn line is given by (\ref{eq:vD}) 
with $D_0=0.4$ and $\langle r^2\rangle=0.27$. The inset shows the effective 
growth rate $v_i$, compared to the theoretical value $x(1-x^2)$ (drawn line). 
The error bars indicate 80\% confidence intervals. 
\label{fig:plotsim}}
\end{figure}


We then perform the fitting procedure outlined above to obtain the tridiagonal 
transition matrix $M$, and in the right panel of Figure~\ref{fig:figsim} we 
have plotted $\tilde{T}=\exp(\tau M)$. There is a clear similarity between the 
two matrices. Figure~\ref{fig:plotsim} shows the resulting values for the 
diffusion coefficient $D$ and the effective growth rate $v$ (inset), computed 
with the help of (\ref{eq:viDi}). The error bars are 80\% confidence intervals 
computed with the bootstrap method (Newman and Barkema, 1999). This captures 
the statistical errors, but not the systematic errors.


\subsection{Comparison with theory}
To place these results in perspective, we compute the Fokker-Planck equation 
for the probability density of $x$ by integrating Eqs.~(5) and (6) of HD over 
the overtone amplitude $r$, to find
\begin{equation}
\frac{\partial \rho}{\partial t}\,=\,
  -\,\frac{\partial}{\partial x}\,x(1-x^2)\rho\,
  +\,\frac{1}{2}\,\frac{\partial^2}{\partial x^2}\,
  D_0(x^2+\langle r^2\rangle|_x)\rho\ .
\label{eq:FP2}
\end{equation}
Hence, we recover Eq.~(\ref{eq:FP1}) with
\begin{equation}
v=x(1-x^2)\ ;\qquad D=D_0(x^2+\langle r^2\rangle|_x)\ .
\label{eq:vD}
\end{equation}
Here $D_0$ a constant equal to $0.4$ for the HD dataset used here, and 
$\langle r^2\rangle|_x$ is the mean square overtone amplitude for given 
$x$. The result is the two drawn lines in Figure~\ref{fig:plotsim}. Since 
$\langle r^2\rangle|_x$ is only a weak function of $x$, we did not bother 
to measure it from the simulation data. Instead, we replaced it by the 
average of $r^2$ over all $x$, measured to be $\langle r^2\rangle=0.27$. 
The $v$ and $D$ recovered from the data compare rather well with their 
theoretical values (\ref{eq:vD}). The agreement for $D$ could be further 
improved by allowing for the fact that $\langle r^2\rangle|_x$ is smaller 
than $0.27$ near $x=0$ and larger than $0.27$ for $x>1$. However, we 
cannot expect agreement to within the statistical errors because of 
approximations made in deriving the Fokker-Planck equation (\ref{eq:FP2}).
As a result there are small systematic differences between the statistical 
properties of $x(t)$ predicted by (\ref{eq:FP2}) and (\ref{eq:vD}) and those 
of the numerically generated $x(t)$. These differences are visible because 
we use many data points ($5\times 10^6$).  

These results demonstrate that our analysis is capable to extract the 
information on the effective VADM growth rate $v$ and the type of noise that 
was used to generate the time series. The scaling $D\propto x^2$ is a 
consequence of the multiplicative noise that the HD model employs [that is, a 
noise term of the type $\dot x=\cdots + N(t)x$]. 
%
But that is really a detail here. The main issue is that we have succesfully 
validated our retrieval method, as we have shown that our analysis is able to 
get out what has been put into the model. To avoid misunderstanding we note 
that this agreement does not say anything on whether the HD model describes 
the physics of the geomagnetic dipole correctly or not, or better than other 
reversal models do. It only tells us that our retrieval method appears to work 
satisfactorily.


\section{Application to the Sint-2000 data}
\begin{figure}[t]
\centerline{\includegraphics[width=6.cm]{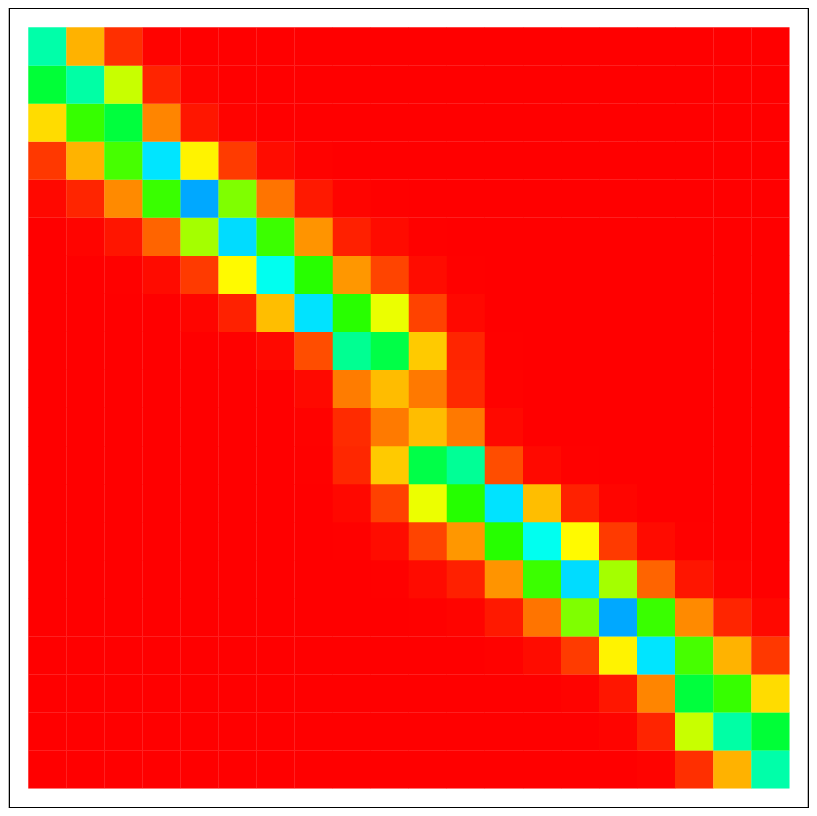}
            \includegraphics[width=6.cm]{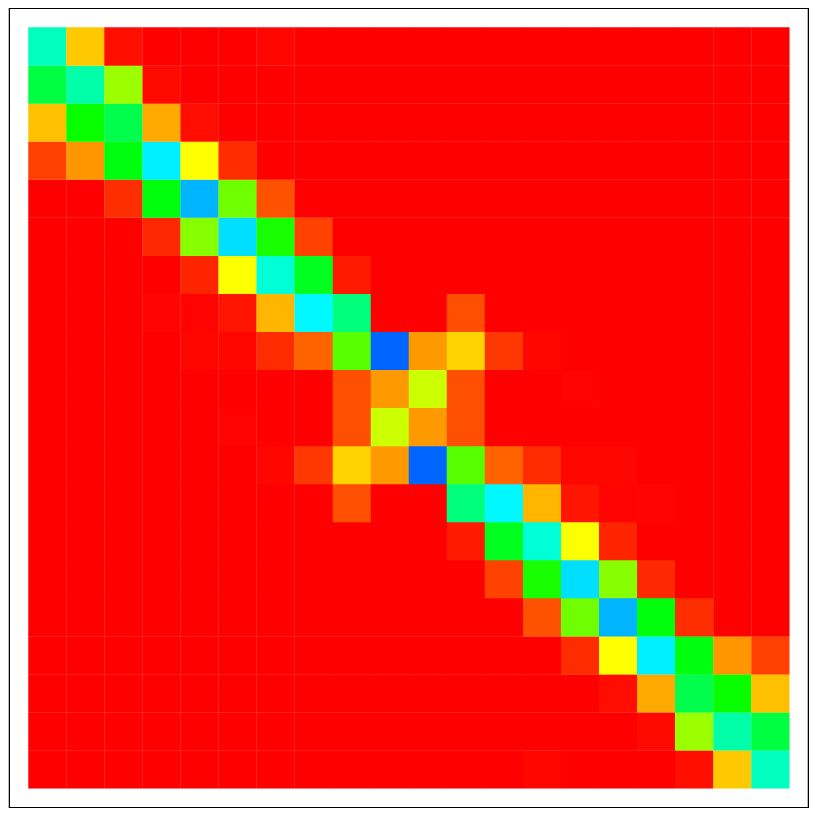}}
\centerline{\includegraphics[width=21.4cm]{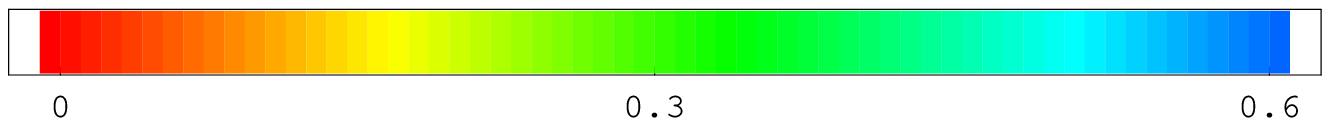}}
\caption{\small Left panel: the transition matrix $T$ obtained from the 
Sint-2000 data, with $\tau=4$ kyr. Right panel: the approximate transition 
matrix $\tilde{T}=\exp(M\tau)$ obtained from a tridiagonal matrix $M$. The 
upper left corner corresponds to $(-10,-10)\cdot 10^{22}$Am$^2$, and the 
lower right corner to $(10,10)\cdot 10^{22}$Am$^2$. We use square bins of 
linear size $1\cdot 10^{22}$Am$^2$.
\label{fig:figsint}}
\end{figure}

We then repeat the same procedure on the Sint-2000 data, 
Figure~\ref{fig:sint2000}. The fitting procedure was performed with a time 
lag of $\tau=4$ kyr, see Figure~\ref{fig:figsint}. This choice is motivated 
as follows. The autocorrelation time of VADM data is a few hunderd yr (the 
time scale for rapid random changes in the geomagnetic dipole), but the 
sampling of the Sint-2000 VADM data increases that to 1 kyr. The time scale 
for systematic changes may be identified with the linear growth time of the 
dipole mode (of the order of 10 kyr). The resulting effective growth rate $v$ 
and diffusion coefficient $D$ are shown in Figure~\ref{fig:plotsint}. The 
error bars are again 80\% confidence intervals computed with the bootstrap 
method (Newman and Barkema, 1999). In reality, the errors will be larger as we 
did not allow for the considerable intrinsic errors in the VADM data (Valet et 
al., 2005).

The $x$ dependence of the effective growth rate is approximately as expected. 
The best fit of the function $\lambda x[1-(x/x_0)^2]$ to the `data points' 
$v_i$ yields $1/\lambda=20^{+13}_{-7}$ ky, and $x_0=(5.4\pm 0.5)\times 
10^{22}\;$Am$^2$. For small $x$ we have $v\propto x$ which corresponds to 
linear growth of the dipole mode when it is small, and the $-x^3$ term is the 
nonlinear quenching. The surprise is in the $x$-dependence of the diffusion 
coefficient $D$ which we discuss below.


\subsection{Implications for the geodynamo}
The analysis of the Sint-2000 data confirms that the geomagnetic dipole mode 
is unstable with a linear growth time $1/\lambda\simeq 20^{+13}_{-7}$ ky. The 
nonlinear quenching follows approximately a quadratic quenching function 
$[1-(x/x_0)^2]$. The nonlinear equilibrium is attained at a VADM of $x_0=5.4
\cdot 10^{22}$Am$^2$. These results are more or less as expected. To our 
knowledge this is the first time that the linear growth rate $\lambda$ and 
the shape of the quenching function of the geomagnetic dipole have been 
measured from pertinent data.

In order to judge our results on the diffusion coefficient we derive the 
theoretical $x$-dependence of $D$. To this end we consider the induction 
equation of MHD: $\pa\vB/\pa t= \na\times(\vv_0\times\vB)+\na\times(\delta\vv
\times\vB)+\eta\na^2\vB$. The fluctuating velocity $\delta\vv$ represents the 
convective turbulence in the metallic outer core superposed on a steady flow 
$\vv_0$. If we expand the magnetic field in the induction equation in 
multipoles, the equation for the dipole becomes $\dot x=\cdots\,+\,{\rm const}
\cdot\delta v(t)x$. Only the contribution of the fluctuating term acting on 
the dipole is written down explicitly. Comparing with Eq.~(\ref{eq:langevin1}) 
which produces a diffusion coefficient $D=2F^2\int_0^\infty\langle 
L(t)L(t-\tau)\rangle d\tau$, we now obtain $D\,\propto\,
(\delta v)_{\rm r.m.s.}^2\tau_{\rm c}\,x^2\,\propto\,\beta\,x^2$, where 
$\tau_{\rm c}$ is the correlation time of $\delta v(t)$, and $\beta\simeq
(\delta v)_{\rm r.m.s.}^2\tau_{\rm c}$ the turbulent diffusion coefficient 
that occurs in the dynamo equation (Moffatt, 1978). Detailed considerations 
lead to (Hoyng, 2008) 
\begin{equation}
D\;\simeq\,\frac{\beta}{R^2}\,\frac{x^2}{N}\;+\;{\rm const}\ ,
\label{eq:Dtype}
\end{equation}
where $R$ is the radius of the outer core and $N$ the number of convective 
cells in the core. There is a small, approximately constant contribution to 
$D$ due to feedback of the overtones on the dipole amplitude. This term also 
occurred in the HD model, cf. Eq.~(\ref{eq:vD}). It is important because it 
is related to the occurrence of reversals but it plays no role in the 
following discussion.

\begin{figure}[t]
\centerline{\includegraphics[width=10.cm]{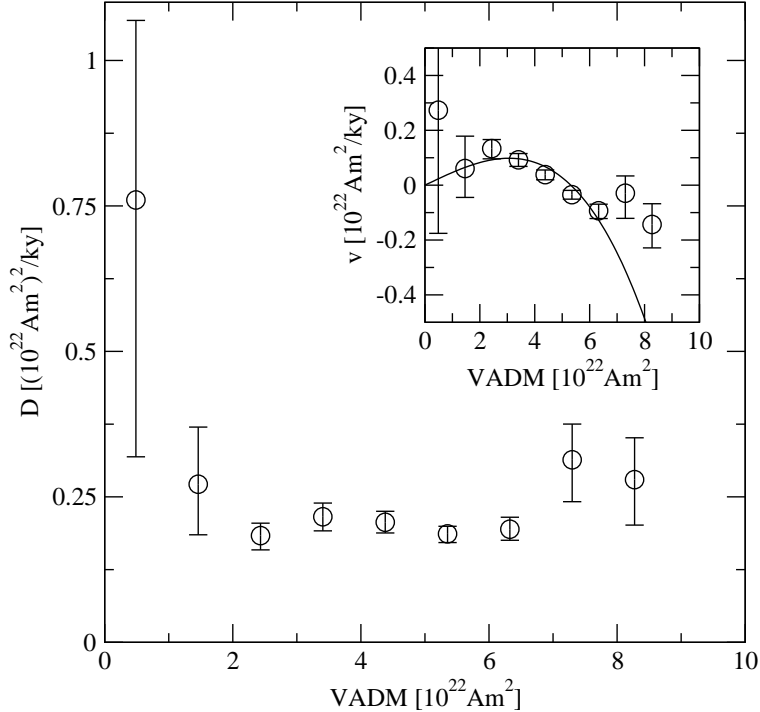}}
\caption{\small Diffusion coefficients $D_i$ and velocity $v_i$ (inset) 
as a function of magnetic dipole strength, obtained from fitting the
Fokker-Planck equation to the Sint-2000 data. The drawn line in the inset
is the best fit of $\lambda x[1-(x/x_0)^2]$ to the `data points' $v_i$, 
see text for details.
\label{fig:plotsint}}
\end{figure}

We expect therefore that $D\propto x^2$, and the explanation is simple. The 
form of the induction equation makes that a given $\delta v$ generates a 
change in $\vB$ proportional to the magnitude of $\vB$. For given $\delta v$ 
the diffusive motion of $\vB$ is therefore larger if $\vB$ is large, and this 
translates to the dipole component as well. However, these considerations are 
not borne out by our numerical results in Figure~\ref{fig:plotsint}.

The increase of the diffusion coefficient for $|x|\rightarrow 0$ in 
Figure~\ref{fig:plotsint} is probably an artifact of the restricted length of 
the data, in combination with the fact that there are only five reversals and 
one aborted reversal in the last 2 Myr. VADMs smaller than $2\times 10^{22}$ 
Am$^2$ are absent in the Sint-2000 data except during the very brief reversal 
periods. Since the data cannot resolve the fine structure of the VADM during 
a reversal one might wonder what the effect would be of a few rapid sign 
changes near a reversal, but that would only serve to make $D(0)$ larger.
  
For VADM $>2\times 10^{22}$ Am$^2$ the diffusion coefficient is constant.
In fact, one might say that the data are consistent with a constant $D$ at
all VADM. We have tested the possibility that this result might somehow be 
caused by the limited time resolution of the Sint-2000 data. To this end we 
have generated from the HD data a Sint-2000-like series by taking a running 
average and then a subset of 2000 data points separated by 1000 yr. This can 
only be done approximately as we cannot convert the dimensionless time of 
the HD data (in units of the dipole growth time) into real time. It is 
conceivable that this new time series would have a large $D(0)$ and a 
constant $D$ at larger $x$, but the resulting $v$ and $D$ did not differ 
materially from those in Figure~\ref{fig:plotsim}. 
 

There seems to be no signature of multiplicative noise in the Sint-2000 data, 
and we believe that this is a solid result. Instead, the data indicate that 
the noise is quasi-additive, because the diffusion coefficient $D\simeq 
F^2L_{\rm r.m.s.}^2\tauc$ in Eq.~(\ref{eq:FP1}) would be constant if $F$ in 
Eq.~(\ref{eq:langevin1}) is constant. We have no explanation for this, but 
there is the intriguing possibility that it is due to a nonlinear quenching 
of the fluid velocity fluctuations $\delta\vv(t)$. If $\beta\propto\langle
(\delta v)^2\rangle\tau_{\rm c}$ would scale as $\propto 1/B^2\propto 1/x^2$, 
then $D$ would be effectively independent of VADM, cf.~(\ref{eq:Dtype}). The 
best way to test these ideas would be to use a longer dataset, and an obvious 
possibility would be to use VADM data from hydromagnetic geodynamo simulations 
for this purpose.


\section{Summary and conclusions}
We have presented and validated a technique for extracting the effective 
growth rate and diffusion coefficient of a time series of a stochastic 
process. An attractive feature of the method is that it does not assume any 
a priori mathematical form for these quantities. Application of the method to 
the Sint-2000 VADM time series has shown that it is possible to measure the 
linear growth rate of the geomagnetic dipole and the shape of the nonlinear 
quenching of this growth rate. The dependence of the diffusion coefficient
on the VADM suggests that the amplitude of the convective flows in the outer 
core is suppressed with increasing dipole strength. The main limitation in 
extracting more useful information on the geodynamo is the length of the 
Sint-2000 series. 
%
%
In future research, we will apply this analysis technique to time series 
obtained from hydromagnetic geodynamo simulations. If no nonlinear quenching 
is observed in these simulations, the simulation model produces time series 
which are qualitatively different from the Sint-2000 VADM time series. On 
the other hand, if these simulations show similar nonlinear quenching, then 
the cause of it can be investigated within the model.


\section*{Acknowledgements}
\label{acknow}
We thank Dr. J.-P. Valet for making the Sint-2000 data available to us, and 
Prof. H. van Beijeren for helpful discussions. 


\section*{References}
\baselineskip=4.mm
{\small

Bouligand, C., Hulot, G., Khokhlov, A. and Glatzmaier, G.A., 2005. Statistical 
paleomagnetic field modelling and dynamo numerical simulation. Geophys. J. 
Int. 161, 603-626.  
\pr
Christensen U., Olson, P. and Glatzmaier, G.A., 1999. Numerical modelling 
of the geodynamo: a systematic parameter study. Geophys. J. Int. 138, 
393-409.
\pr
Constable, C.G. and Parker, R.L., 1988. Statistics of the geomagnetic secular 
variation in the past 5 Myr. J. Geophys. Res. 93, 11569-11581.
\pr
Gardiner, C.W., 1990. Handbook of Stochastic Methods. Springer-Verlag, 
Berlin.
\pr
Glatzmaier, G.A. and Roberts, P.H., 1995. A three-dimensional 
self-consistent computer simulation of a geomagnetic field reversal.
Nature 377, 203-209.
\pr
Guyodo, Y. and Valet, J.-P., 1999. Global changes in intensity of the 
Earth's magnetic field during the past 800 kyr. Nature 399, 249-252.
\pr
Hoyng, P., Schmitt, D. and Ossendrijver, M.A.J.H., 2002. A theoretical 
analysis of the observed variability of the geomagnetic dipole field. 
Phys. Earth Planet. Int. 130, 143-157.
\pr
Hoyng, P. and Duistermaat, J.J., 2004. Geomagnetic reversals and the 
stochastic exit problem. Europhys. Lett. 68, 177-183. (HD)
\pr
Hoyng, P., 2008. To be submitted.
\pr 
Hulot, G. and Le Mou\"el, J.L., 1994, A statistical approach to the 
Earth's main magnetic field. Phys. Earth Planet. Int. 82, 167-183.
\pr
Kuang, W. and Bloxham, J., 1997. An Earth-like numerical dynamo model.
Nature 389, 371-374.
\pr
Merrill, R.T., McElhinny, M.W. and McFadden, P.L., 1996. The Magnetic
Field of the Earth. Academic Press, New York.
\pr
Moffatt, H.K., 1978. Magnetic Field Generation in Electrically Conducting 
Fluids. Cambridge U.P.
\pr
Newman, M.E.J. and Barkema, G.T., 1999. Monte Carlo Methods in Statistical 
Physics. Oxford U.P.
\pr
Valet, J.-P., Meynadier, L. and Guyodo, Y., 2005. Geomagnetic dipole 
strength and reversal rate over the past two million years. Nature 435, 
802-805.
\pr
Van Kampen, N.G., 1992. Stochastic Methods in Physics and Chemistry. 
North-Holland, Amsterdam. 
}


\end{document}